\documentclass[aps,prb,twocolumn,superscriptaddress,showpacs]{revtex4-1}

\usepackage{graphicx}
\usepackage{amsmath}

\newcommand{\rvo}{\textit{R}VO$_3$}
\newcommand{\smvo}{SmVO$_3$}
\newcommand{\mono}{$P2_1/b$}
\newcommand{\orth}{$Pbnm$}

\newcommand{\etal}{\textit{et al.}}
\newcommand{\tso}{$T_{\mathrm{N}}$}
\newcommand{\tsol}{$T_{\mathrm{N}2}$}
\newcommand{\too}{$T_{\mathrm{OO}}$}
\newcommand{\tool}{$T_{\mathrm{OO}2}$}

\begin{document}

\title{An x-ray diffraction study of the temperature-induced structural phase transitions in \smvo}

\author{R. D. Johnson}\email{r.johnson1@physics.ox.ac.uk}
\affiliation{Clarendon Laboratory, Department of Physics, University of Oxford, Oxford, OX1 3PU, United Kingdom}
\affiliation{ISIS facility, Rutherford Appleton Laboratory-STFC, Chilton, Didcot, OX11 0QX, United Kingdom}
\author{C. C. Tang}
\affiliation{Diamond Light Source, Harwell Science and Innovation Campus, Didcot, OX11 0DE, United Kingdom}
\author{I. R. Evans}
\affiliation{Department of Chemistry, Durham University, South Road, Durham, DH1 3LE, United Kingdom}
\author{S. R. Bland}
\affiliation{Department of Physics, Durham University, South Road, Durham, DH1 3LE, United Kingdom}
\author{D. G. Free}
\affiliation{Department of Chemistry, Durham University, South Road, Durham, DH1 3LE, United Kingdom}
\author{T. A. W. Beale}
\author{P. D. Hatton}
\affiliation{Department of Physics, Durham University, South Road, Durham, DH1 3LE, United Kingdom}
\author{L. Bouchenoire}
\affiliation{European Synchrotron Radiation Facility, B. P. 220, F-38043 Grenoble Cedex, France}
\author{D. Prabhakaran}
\author{A. T. Boothroyd}
\affiliation{Clarendon Laboratory, Department of Physics, University of Oxford, Oxford, OX1 3PU, United Kingdom}

\date{\today}

\begin{abstract}

Through powder x-ray diffraction we have investigated the structural behavior of \smvo, in which orbital and magnetic degrees of freedom are believed to be closely coupled to the crystal lattice. We have found, contrary to previous reports, that \smvo\ exists in a single, monoclinic, phase below 200~K. The associated crystallographic distortion is then stabilized through the magnetostriction that occurs below 134~K. The crystal structure has been refined using synchrotron x-ray powder diffraction data measured throughout the structural phase diagram, showing a substantial Jahn-Teller distortion of the VO$_6$ octahedra in the monoclinic phase, compatible with the expected G-type orbital order. Changes in the vanadium ion crystal field due to the structural and magnetic transitions have then been probed by resonant x-ray diffraction.

\end{abstract}


\maketitle

\section{\label{intro}Introduction}

Many advances in solid state physics have been led by research into the interaction of spin, orbital and charge degrees of freedom within single crystals\cite{tokura03}. These microscopic ordering processes, strongly coupled to the crystal lattice, are particularly evident in transition metal oxides. In some such materials, it has been shown that it is possible to manipulate the electronic correlations by the control of external parameters. For example, colossal magneto-resistance was discovered in thin film La$_{0.67}$Ca$_{0.33}$MnO$_x$, in which at 77 K the application of a 6 Tesla magnetic field induced more than a thousand-fold drop in resistivity (corresponding to a huge magnetoresistance ratio of 127,000 \%)\cite{jin94}. Also, in multiferroic TbMn$_2$O$_5$, a complete reversal of the electric polarization is induced by an applied magnetic field of 2 Tesla\cite{hur04,johnson11}. The understanding of such ordering phenomena in these systems is therefore of fundamental scientific interest and potentially of great technological benefit.

\begin{figure}
\includegraphics[width=8cm]{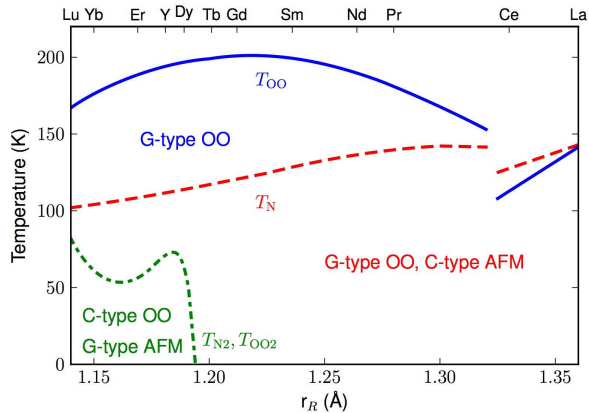}
\caption{\label{pdfig}(Color online) The \rvo\ series phase diagram redrawn from Miyasaka \etal\cite{miyasaka03} \too, \tso, and \tool/\tsol\ are marked by the blue solid, red dashed and green dash dotted lines, respectively.}
\end{figure}

The distorted perovskite \rvo\ series (\textit{R} = rare-earth ion or yttrium) has attracted substantial and sustained study\cite{miyasaka03,yan04,tung07,nguyen95,horsch08} as they form a set of materials in which orbital, spin and lattice degrees of freedom are believed to be closely coupled. The \rvo\ phase diagram determined by Miyasaka \etal\cite{miyasaka03} (figure \ref{pdfig}) splits the series into three subgroups. The small rare-earth radius (\textit{R} = Lu to Dy) compounds undergo three successive phase transitions, firstly developing a \textit{G}-type orbital order (OO) of the V$^{3+}$ $3d$ states at \too\ $\sim 170 - 200$ K, and then an accompanying \textit{C}-type antiferromagnetic order (AFM), also involving the V$^{3+}$ $3d$ electrons, at \tso\ $\sim 90 - 120$ K. Unique to this subgroup is a further transition at \tool\ = \tsol\ $\sim 50 - 80$ K, at which the \textit{G}-type OO and \textit{C}-type AFM switch to \textit{C}-type OO and \textit{G}-type AFM. By comparison, the large rare-earth radius (\textit{R} = La and Ce) compounds exhibit just two ordered phases. Unusually, the \textit{C}-type AFM evolves first at \tso\ $\sim 110 - 140$ K, with \too\ occurring a few degrees lower. The remaining middle members of the series (\textit{R} = Tb to Pr) form the third subgroup that enter the \textit{G}-type OO phase below \too\ $\sim 180 - 200$ K with \textit{C}-type AFM developing below \tso\ $\sim 110 -140$ K. The orbital phase transitions are coincident with structural phase transitions; evidence of the strong coupling to the crystal lattice\cite{yan04}. All members of the \rvo\ series are orthorhombic at room temperature adopting the space group \orth. At \too\ the crystal symmetry is then lowered to the monoclinic space group, \mono. Further, the first subgroup of compounds returns to \orth\ symmetry upon cooling below \tool\ [\citenum{miyasaka03}].

In this paper we focus on the \textit{R} = Sm member of the series belonging to the third subgroup outlined above. In \smvo, heat capacity measurements show \too\ = 192.6~K and \tso\ = 130~K [\citenum{tung07}]. The majority of \rvo\ compounds (excluding \textit{R} = Pr, Ce, Lu) exhibit magnetization reversal (MR). Upon cooling \smvo\ below \tso, the magnetization, which is initially positive, smoothly reverses twice at \textit{T} = 127.5 and 63.8 K [\citenum{tung07}]. A number of mechanisms for the MR have been proposed\cite{kimishima00,ren98,tung07}. Most recently Tung \etal\cite{tung07} explained that the MR occurs due to a minority of random field spins, forming a separate magnetic sublattice from the long-range, strongly coupled antiferromagnetic order. The random field spins are then believed to originate from defects in the orbital system. Assuming this model to be correct, the strong coupling of the orbital degrees of freedom to the lattice is therefore important in explaining the MR observed.

In a high-resolution x-ray powder diffraction study by Sage \etal\cite{sage06,sage07} it was proposed that, just below \tso, \smvo\ has coexisting orbital and structural phases. In this scenario the orbital degrees of freedom are strongly coupled to the symmetry of the crystal lattice. It was suggested that orthorhombic, \textit{C}-type orbitally ordered droplets form in the larger monoclinic, \textit{G}-type orbitally ordered crystal, stabilized by strain at the crystal/droplet boundaries. This could result from the competition between the extreme octahedral tilting (promoting \textit{C}-type OO) observed in the smaller radius rare-earth compounds and the reduction of the unit cell volume by magnetostriction (promoting \textit{G}-type OO) that is significant in the larger radius rare-earth compounds. This phase separation is therefore possibly common to Tb, Gd, Eu and Sm members of the \rvo\ series due to their similar, middle rare-earth radius. Furthermore, Tung \etal\cite{tung07} commented that the phase separation is consistent with, but not equivalent to, their model of MR based upon the existence of random field spins. However, in a neutron diffraction study\cite{reehuis06}, no phase separation was observed in single crystal TbVO$_3$. This was then addressed by Sage \etal\cite{sage06,sage07} who suggested that in the single crystal sample insufficient strain at the crystal/droplet boundaries can form, hence making the phase separation unstable.

Here we report an investigation of the structural phases of \smvo\ via two x-ray diffraction techniques. Typically, the crystal structure of transition metal oxides would be determined by neutron diffraction. However, samarium has a very high neutron absorption cross-section ($\sigma_\mathrm{a} =$ 5922 barn for 2200 m s$^{-1}$ neutrons) making neutron diffraction unsuitable. Indeed, the Sm, Eu, Gd and Dy members of the \rvo\ series were omitted from a thorough crystallographic survey\cite{martinez-lope08} for this reason. We have determined the crystalline symmetry and refined the respective lattice parameters of \smvo\ from lab-based x-ray powder diffraction data measured in the temperature range 12 K to 300 K. The crystal structure was then refined against synchrotron x-ray powder diffraction data measured in all structural phases. Furthermore, we have investigated the crystal field local to the vanadium and samarium ions by resonant x-ray diffraction (RXD). From the results presented we discuss the coupling of the crystal lattice to the orbital and magnetic electronic orders.

\section{\label{exp}Experiment}

A single crystal sample of \smvo\ was grown by the floating zone technique\cite{prabhak03,prabhak05}. A powder sample was prepared by grinding an off-cut of the single crystal using an agate pestle and mortar. Prior to x-ray diffraction measurements, the sample magnetization was characterized as a function of temperature using a Quantum Design SQUID magnetometer.

Laboratory-based variable-temperature powder x-ray diffraction data were collected using a Bruker D8 diffractometer, with a LynxEye Si strip detector and an Oxford Cryosystem PheniX CCR cryostat. An internal silicon standard was used for accurate unit cell determination. The sample was cooled and warmed between 12 and 300~K at a rate of 10 K hr$^{-1}$. Data were collected over a 2$\theta$ range of 15-120$^\circ$ for 30 minutes, giving an average of one scan every 5 K. Unit cell parameters as a function of temperature were extracted from data analysis by two-phase Rietveld fitting (\smvo\ and Si) using the Bruker DIFFRACplus TOPAS software package\cite{topas}. A monoclinic structural model for \smvo\ in space group $P2_1/b11$ was used throughout the temperature range observed, so that the alpha angle was allowed to freely refine away from 90 degrees.

High resolution synchrotron x-ray powder diffraction measurements were performed at beam line I11, Diamond light source\cite{tang07,thompson09}. The powder sample was adhered to the outer surface of a glass capillary in order to minimise absorption. Data were collected at 105, 165, and 295 K over a 2$\theta$ range of 10-100$^\circ$ with a x-ray wavelength of 0.82615 $\mathrm{\AA}$, against which the crystal structure of \smvo\ was refined using the FullProf suite of programs\cite{rodriguezcarvaja93}.

Resonant x-ray diffraction (RXD) measurements were performed at the XMaS UK CRG beam line\cite{brown01} at the ESRF. A single crystal sample of approximate size 2 x 2 x 2 mm$^3$ was prepared with (011) and (010) surface normal facets, cut and polished to a roughness of 1 micron. The sample was mounted on the cold finger of a closed cycle helium cryostat fitted to a six-circle diffractometer. The crystal field local to the vanadium ions was probed by tuning to the resonant enhancement of anisotropic tensor of susceptibility\cite{dmitrienko83} (ATS) reflections that exists at ion specific absorption edges. The sample was mounted with the (022) Bragg reflection surface normal. The (011) Bragg forbidden, ATS reflection was then located. The ATS scattered intensity was measured whilst scanning the incident x-ray energy through the vanadium \textit{K}-edge, hence measuring the crystal field local to the vanadium sites. Energy spectra were measured at 105 K, 165 K and 295 K, corresponding to the three phases of \smvo. This was then repeated for the (010) ATS reflection, having remounted the sample with the (020) Bragg reflection surface normal. By measuring the energy spectra at a number of different azimuth angles, we were able to identify and exclude contamination from multiple scattering processes. All measurements were made in the rotated $\sigma-\pi'$ channel, employing a pyrolytic graphite polarization analyser crystal scattering at the (004) reflection.

\section{\label{randd}Results and discussion}

\begin{figure}
\includegraphics[width=8cm]{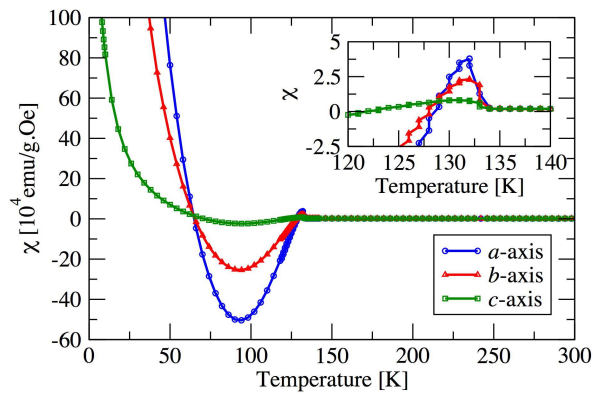}
\caption{\label{magfig}(Color online) Mass magnetic susceptibility of \smvo\ measured along the \textit{a-} (blue circles), \textit{b-} (red triangles) and \textit{c-} (green squares)  crystallographic axes. Measurements were made upon cooling from 300 K down to 2 K in a small applied field of 10 Oe. The inset shows the magnetic transition in greater detail.}
\end{figure}

Figure \ref{magfig} shows the mass magnetic susceptibility of \smvo, measured along the three crystallographic axes, as a function of temperature. Data were taken upon field cooling in 10 Oe from 300~K down to 2~K. In this sample, \tso\ was found to be at 134~K, shown in greater detail in the inset. After an initial increase, the magnetization smoothly reverses at 129~K and 64.5~K, below which the susceptibility continues to increase down to the lowest achievable temperature. This trend is in good agreement with that published by Tung \etal\cite{tung07} Interestingly, however, we observe a greater anisotropy. Tung \etal\ showed that the \textit{a}-axis is easy, with lower, approximately equal susceptibilities measured along the \textit{b-} and \textit{c-}axes. In our measurement it is clear that the susceptibility along the \textit{b}-axis is intermediate with respect to the easy \textit{a}-axis and the hard \textit{c}-axis.

\subsection{\label{powdersec}X-ray powder diffraction}

\begin{figure}
\includegraphics[width=8.5cm]{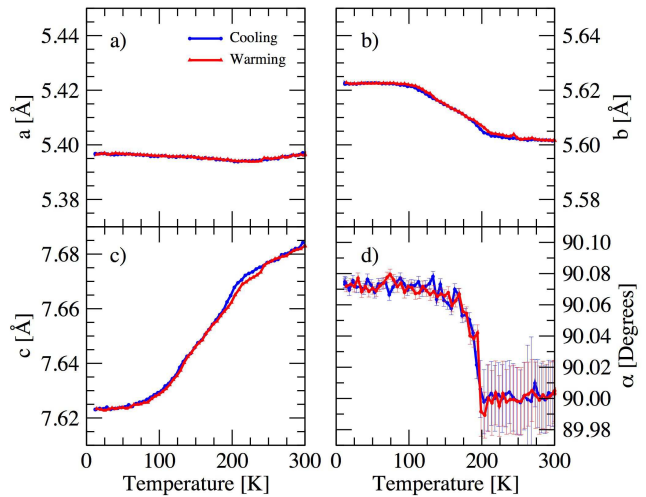}
\caption{\label{lattfig}(Color online) Lattice parameters refined from lab-based x-ray powder diffraction data measured as a function of temperature. The \textit{a, b,} and \textit{c}  lattice parameters are shown on a relative vertical scale.}
\end{figure}

Powder x-ray diffraction was used to monitor the structural changes in \smvo\ between room temperature and 12 K. In order to determine accurately the unit cell parameters of \smvo, a monoclinic structural model was Rietveld-refined against data measured throughout the observed temperature range, allowing the angle $\alpha$ to freely vary from, or refine back to, 90$^\circ$. Figure \ref{lattfig} shows the obtained results for the unit cell parameters \textit{a, b, c} and $\alpha$. Both cooling and warming data are given, showing no thermal hysteresis. The lattice parameter \textit{a} changes very little over the whole temperature range. By comparison, upon \too\ we observed a significant expansion of \textit{b}, which then reaches its maximum value at $\sim$100~K. The cell parameter \textit{c} exhibits the largest change, decreasing over the entire temperature range. Figure \ref{lattfig}d shows the monoclinic angle, $\alpha$, as a function of temperature. The freely refined alpha angle stays at 90$^\circ$ until just below 200 K, at which point the crystal structure starts distorting and the structural phase transition occurs. Below \too, $\alpha$ increases sharply until below \tso, at which point the monoclinic distortion locks into a value of $\sim$90.075(3)$^\circ$. There are no further deviations or any changes in the behaviour of the lattice parameters that suggest additional structural phase changes in the sample at low temperatures. 

\begin{figure}
\includegraphics[width=8.5cm]{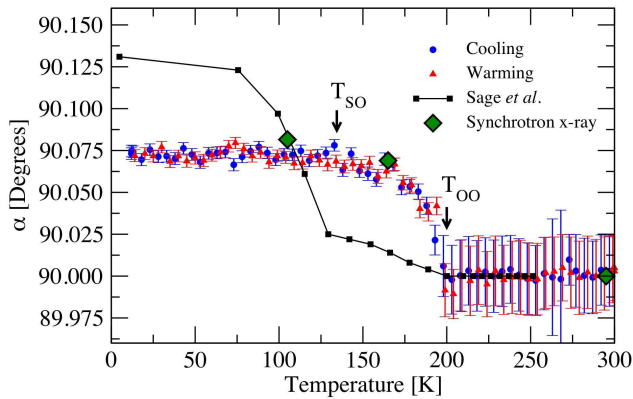}
\caption{\label{alphafig}(Color online) The monoclinic angle, $\alpha$, refined from lab-based x-ray powder diffraction data measured as a function of temperature. The equivalent data measured by Sage \etal\cite{sage06} are reproduced and overlaid (black line). The alpha angles found from synchrotron x-ray diffraction are indicated by green diamonds.}
\end{figure}

Our results differ from those of Sage \etal\cite{sage06}, who observed a re-emergence of the orthorhombic form of \smvo\ below \tso; these different findings do not seem to be a mere consequence of different resolutions of the two instruments used (synchrotron vs. lab source). A closer comparative analysis of the monoclinic distortion trends observed in the two studies (Fig. \ref{alphafig}) seems to suggest a genuinely different behaviour of the two samples. Our sample shows two regions of behaviour: on cooling the orthorhombic structure below \too, the monoclinic distortion starts increasing abruptly and locks into values which persist down to the lowest observed temperature. The $\alpha$ vs. temperature dependence found by Sage \etal\ for their sample, on the other hand, exhibits three distinctive regions: one in which $\alpha$ is constant (room temperature to \too), an intermediate region where it increases very gradually (\too\ to \tso) and a third region characterised by a much sharper increase in $\alpha$ (below \tso). Another important point is that the base temperature values found for the monoclinic angle $\alpha$ in the two samples are quite different, as can be appreciated from the plots in Fig. \ref{alphafig}. One possible explanation for the observed discrepancies is differences in oxygen content, where a deficiency may lead to phase coexistence. Thermogravimetric analysis of our sample gave an oxygen content of $3.08\pm0.05$, indicating high stoichiometric accuracy, and that the structural behavior reported here is intrinsic to \smvo. We surmise that the locking of the monoclinic distortion is due to magnetostriction within the sample. The onset of magnetic order at \tso\ is therefore closely coupled to the distortions of the crystal structure in the low-temperature phase.

\begin{figure}
\includegraphics[width=8cm]{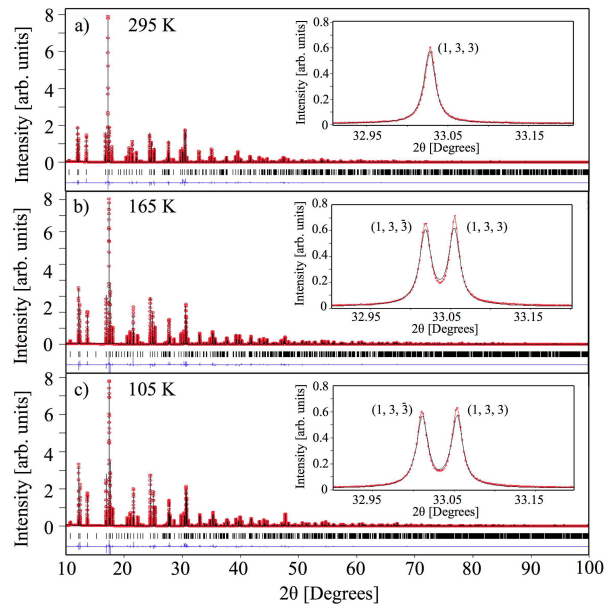}
\caption{\label{patternfig}(Color online) Synchrotron x-ray powder diffractograms measured at a) 295, b) 165, and c) 105 K. Data points and the calculated pattern are shown as red circles and a black curve respectively. Tick marks indicate the position of Bragg reflections according to the respective space groups and lattice parameters. A difference profile (observed - calculated) is shown as a blue curve at the bottom of each pane. Insets: the well resolved splitting of the (1, 3, 3) Bragg peak upon the structural transition form orthorhombic to monoclinic.}
\end{figure}

Given the differences between this study and that of Sage \textit{et al}., we have performed a synchrotron x-ray powder diffraction experiment at I11, Diamond Light Source, in order to accurately determine the crystal structure in all three phases. A powder diffraction experiment was chosen in preference to a single crystal measurement to avoid complications that arise due to pseudo-merohedral twinning at the orthorhombic to monoclinic phase transition. Three data sets were measured at 295, 165 and 105 K, shown in figure \ref{patternfig}. A crystal structure model corresponding to that previously reported\cite{sage06} for \smvo\ was refined against the data, giving an excellent fit with $R_\mathrm{Bragg}$ values 6.95, 5.96, and 6.24 $\%$, respectively. The peak splitting due to the reduction of crystal symmetry from orthorhombic to monoclinic at \too\ was well resolved, as highlighted in the insets of Fig. \ref{patternfig}. The corresponding monoclinic angles, $\alpha$, are marked in figure \ref{alphafig}. They were found to be in good agreement with the temperature dependence of alpha, hence validating the laboratory-based data and supporting the observation of genuinely different behavior between our sample and that of Sage \textit{et al}.

For each phase the structural parameters and a selection of bond lengths and angles are given in Table \ref{structureparams}, with the corresponding anisotropic atomic displacement parameters summarized in Table \ref{thermalparams}.

Orbital order in transition metal oxides is coupled to the crystal lattice through Jahn-Teller (JT) distortions. In \smvo, vanadium ions have valence 3+, corresponding to two spin-up electrons occupying two $t_{2g}$ states. The vanadium ions are located within octahedral oxygen coordinations, in which the crystal electric field splitting of the valence $d$-orbitals gives rise to energetically degenerate $t_{2g}$ states. In \rvo\ it is understood that this degeneracy is lifted through a distortion of the octahedra\cite{miyasaka03} (JT distortion). This structural modulation lengthens two opposite V-O bonds and shortens the others. The direction of the long bonds alternates throughout the crystal, giving an alternating preferential occupation of the $d_{yz}$ and $d_{zx}$ orbitals (whilst the $d_{xy}$ states remain occupied by a single electron).

\begin{figure}
\includegraphics[width=8cm]{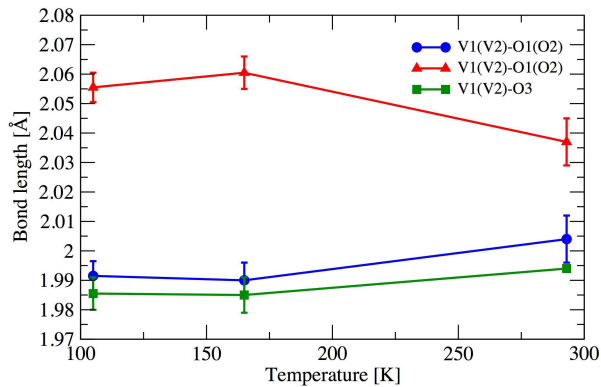}
\caption{\label{bondlengthfig}(Color online) Refined V-O bond lengths showing the Jahn-Teller distortion of the orbitally ordered phases. In the monoclinic phases, the V-O bond lengths equivalent in the orthorhombic phase have been averaged.}
\end{figure}

In \smvo, we have observed the lengthening and shortening of V-O bonds at \too, as predicted. Figure \ref{bondlengthfig} summarizes this result, showing the three non-equivalent bond lengths as a function of temperature. In the monoclinic phase the octahedra centered at ($\tfrac{1}{2}$,$\tfrac{1}{2}$,$\tfrac{1}{2}$) and ($\tfrac{1}{2}$,$\tfrac{1}{2}$,0) become inequivalent. In this analysis, the bond lengths once equivalent in orthorhombic symmetry have therefore been averaged. We note, also, that the refined V-O bond lengths are consistent with the JT distortions associated with G-type orbital order (as opposed to C-type) in both orbitally ordered, low temperature phases.

In the \rvo\ series it has been shown that the G-type orbital order, as observed in \smvo\ below \too, competes with orbital fluctuations and the associated structural disorder\cite{ulrich03,blake09}. Such disorder was even inferred to exist in HoVO$_3$ at room temperature\cite{blake09}. Two important consequences for the orbital behavior of the system follow this argument. First, if strong orbital fluctuations were to occur, the expected cooperative Jahn-Teller distortions would be completely suppressed. Secondly, the degeneracy of the $d_{xz}$/$d_{yz}$ orbital occupation, in which the fluctuations originate, could be lifted by a spontaneous dimerization of the orbitals along the \textit{c}-axis; \textit{i.e.} a 1D orbital Peierls state with an associated small structural distortion. In this structural study the JT distortion is clearly observable, indicating that orbital fluctuations and dimerization are not prevalent in \smvo.

Sage \etal\cite{sage06,sage07} commented that the degree of octahedral tilting (known as the GdFeO$_3$ distortion) is crucial in determining the nature of the orbital order in the \rvo\ series. The large tilting in compounds with small rare-earth radius favours \textit{C}-type orbital order, whereas in compounds with larger rare-earth radius, magnetostriction dominates over a smaller octahedral tilting, favouring \textit{G}-type orbital order.  A crystallographic study on this series, in which \smvo\ was omitted, reports room temperature tilting angles of the extreme members of the series, LuVO$_3$ and LaVO$_3$, to be 19.1$^\circ$ and 11.6$^\circ$, respectively\cite{martinez-lope08}. From our 295 K structure refinement we find a tilting angle of 15.5(1)$^\circ$ for \smvo, calculated from the V1-O3-V1(V2) angle given in Table \ref{structureparams}. This is almost exactly in the centre of the Lu and La tilting angles, supporting the argument of Sage \etal\cite{sage06,sage07} (who measured a tilting angle of 15.2(2)$^\circ$) that the octahedral tilting in \smvo\ is intermediate with respect to the whole series.

In the following section we describe the use of resonant x-ray diffraction to investigate the crystal field distortions local to the vanadium ions, which are key to the splitting of the electronic $d$-states and hence the existence of JT distortions and orbital order.

\begin{table}
\caption{\label{structureparams}Results of the synchrotron x-ray powder diffraction structure refinement of \smvo. The data were collected at 105~K, 165~K and 295~K, and refined in the space groups listed. The atomic fractional coordinates are given, along with selected bond distances (\AA) and angles (degrees).}
\begin{ruledtabular}
\begin{tabular}{l c c c}
& 105 K & 165 K & 295 K \\
\hline
Space group & $P2_1/b$ & $P2_1/b$ & $Pbnm$\\
$x$(Sm) & 0.51488(4) & 0.51455(4) & 0.51372(6) \\
$y$(Sm) & 0.94118(4) & 0.94133(4) & 0.94278(5) \\
$z$(Sm) & 0.75002(4) & 0.75019(5) & 0.75 \\
$x$(V1) & 0.5 & 0.5 & 0.5 \\
$y$(V1) & 0.5 & 0.5 & 0.5 \\
$z$(V1) & 0.5 & 0.5 & 0.5 \\
$x$(V2) & 0.5 & 0.5 & - \\
$y$(V2) & 0.5 & 0.5 & - \\
$z$(V2) & 0 & 0 & - \\
$x$(O1) & 0.202(1) & 0.203(1) & 0.2008(5) \\
$y$(O1) & 0.3031(8) & 0.3038(9) & 0.2999(5) \\
$z$(O1) & 0.4489(7) & 0.4467(7) & 0.4492(4) \\
$x$(O2) & 0.696(1) & 0.6957(9) & - \\
$y$(O2) & 0.2078(8) & 0.2080(9) & - \\
$z$(O2) & 0.9486(6) & 0.9496(6) & - \\
$x$(O3) & 0.5952(5) & 0.5958(6) & 0.5934(7) \\
$y$(O3) & 0.4650(6) & 0.4676(6) & 0.4694(7) \\
$z$(O3) & 0.2510(8) & 0.2491(8) & 0.25 \\
\\
\hline
\\
V1-O1 & 1.990(5)$\times2$ & 1.983(6)$\times2$ & 2.004(8)$\times2$ \\
V1-O1 & 2.061(5)$\times2$ & 2.070(6)$\times2$ & 2.037(8)$\times2$ \\
V1-O3 & 1.978(5)$\times2$ & 1.996(6)$\times2$ & 1.994(1)$\times2$ \\
V2-O2 & 1.993(5)$\times2$ & 1.987(6)$\times2$ & - \\
V2-O2 & 2.050(5)$\times2$ & 2.051(5)$\times2$ & - \\
V2-O3 & 1.993(6)$\times2$ & 1.984(6)$\times2$ & - \\
\\
V1-O1-V1 & 148.2(2) & 147.7(2) & 148.5(9) \\
V2-O2-V2 & 149.0(2) & 149.1(2) & - \\
V1-O3-V1(V2) & 147.8(2) & 148.0(2) & 149.0(1) \\
\end{tabular}
\end{ruledtabular}
\end{table}

\begin{table}
\caption{\label{thermalparams}Thermal parameters for \smvo\ for the structures refined in all three phases ($\times100$~\AA$^2$).}
\begin{ruledtabular}
\begin{tabular}{l c c c}
& 105 K & 165 K & 295 K \\
\hline
$U_{11}$(Sm) & 1.725(10) & 1.763(10) & 1.968(11) \\
$U_{22}$(Sm) & 1.603(8) & 1.675(9) & 1.848(10) \\
$U_{33}$(Sm) & 1.584(9) & 1.647(9) & 1.821(9) \\
$U_{12}$(Sm) & -0.038(8) & -0.028(9) & -0.103(14) \\
$U_{13}$(Sm) & -0.073(10) & -0.076(12) & 0 \\
$U_{23}$(Sm) & -0.040(18) & 0.03(2) & 0 \\
\\
$U_{11}$(V1) & 1.55(8) & 1.58(8) & 1.90(4) \\
$U_{22}$(V1) & 1.73(6) & 1.47(6) & 1.74(3) \\
$U_{33}$(V1) & 1.69(6) & 1.76(6) & 1.81(3) \\
$U_{12}$(V1) & -0.04(4) & 0.02(4) & -0.03(4) \\
$U_{13}$(V1) & 0.02(8) & -0.03(9) & -0.02(4) \\
$U_{23}$(V1) & 0.09(6) & -0.07(7) & -0.02(2) \\
\\
$U_{11}$(V2) & 1.65(8) & 1.77(9) & - \\
$U_{22}$(V2) & 1.54(6) & 1.72(7) & - \\
$U_{33}$(V2) & 1.76(6) & 1.62(6) & - \\
$U_{12}$(V2) & -0.07(4) & -0.03(4) & - \\
$U_{13}$(V2) & 0.03(8) & 0.09(9) & - \\
$U_{23}$(V2) & 0.01(6) & -0.05(8) & - \\
\\
$U_{11}$(O1) & 1.7(3) & 4.1(4) & 1.92(13) \\  
$U_{22}$(O1) & 2.0(2) & 1.3(2) & 2.46(15) \\
$U_{33}$(O1) & 3.0(3) & 2.9(3) & 3.11(16) \\
$U_{12}$(O1) & -0.01(18) & 0.01(2) & -0.06(11) \\
$U_{13}$(O1) & 0.07(18) & 0.02(2) & -0.73(11) \\
$U_{23}$(O1) & -0.4(2) & -0.16(3) & 0.64(12) \\
\\
$U_{11}$(O2) & 1.80(14) & 2.37(16) & - \\
$U_{22}$(O2) & 2.01(19) & 1.82(20) & - \\
$U_{33}$(O2) & 1.83(16) & 1.88(17) & - \\
$U_{12}$(O2) & -0.42(18) & -0.19(19) & - \\
$U_{13}$(O2) & 0.68(13) & 0.54(13) & - \\
$U_{23}$(O2) & 0.2(2) & 0.2(2) & - \\
\\
$U_{11}$(O3) & 1.0(3) & 0.3(18) & 2.13(17) \\
$U_{22}$(O3) & 2.3(3) & 2.6(3) & 3.2(2) \\
$U_{33}$(O3) & 2.7(3) & 2.3(3) & 1.97(17) \\
$U_{12}$(O3) & -0.52(18) & -0.24(18) & 0.41(15) \\
$U_{13}$(O3) & 0.18(17) & -0.08(15) & 0 \\
$U_{23}$(O3) & -0.41(18) & 0.1(16) & 0
\end{tabular}
\end{ruledtabular}
\end{table}

\begin{figure}
\includegraphics[width=8 cm]{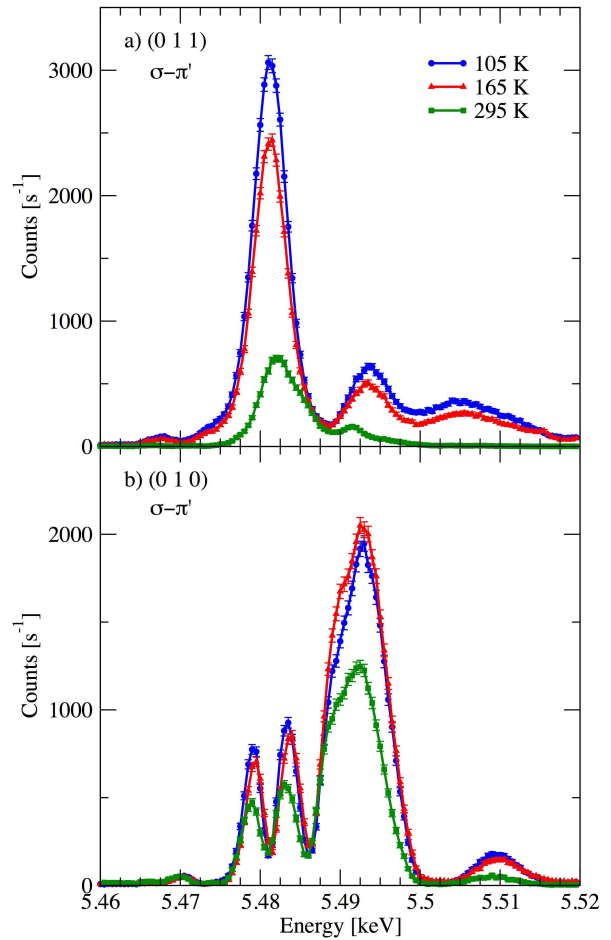}
\caption{\label{resfig}(Color online) Scans of energy at fixed a) (011) and b) (010) wavevector, measured through the vanadium \textit{K}-edge. Data was taken at 105 K (blue), 165 K (red) and 295 K (green), corresponding to the different phases of \smvo.}
\end{figure}

\subsection{\label{rxssec}Resonant x-ray diffraction}

Incident x-rays were tuned to the vanadium \textit{K}-edge exciting a resonant transition between the bound vanadium $1s$ electronic states and the delocalized vanadium $4p$ continuum of states. By measuring anisotropic tensor of susceptibility (ATS)\cite{dmitrienko83} diffraction signals at Bragg-forbidden reciprocal space points, we are sensitive to the crystal field local to the vanadium ions. However, the origin of reflections at Bragg-forbidden positions is often difficult to assign, as a number of long-range electronic orders (\textit{e.g}. magnetic) can be present at the same wavevector. Indeed, in a previous study of YVO$_3$\cite{noguchi00}, Bragg-forbidden reflections were considered to be direct evidence for orbital order. In this scenario the origin of the anisotropy of the $4p$ states, ultimately resulting in the resonant diffraction signal, lies in the $3d$ - $4p$ Coulomb interaction. The diffraction measurement would therefore be sensitive to any long-range preferential occupation of the $3d$ orbitals. However, the contribution of this effect to the \textit{K}-edge resonant diffraction signal is small. In orbitally ordered LaMnO$_3$, the Mn \textit{K}-edge diffraction signal due to the Coulomb interaction was shown (via \textit{ab-initio} calculations) to be 100 times weaker than that originating in the $4p$ anisotropy induced by the distorted oxygen octahedra that surround the manganese ions\cite{benfatto99}. Furthermore, in YVO$_3$, theoretical predictions showed that the observed resonant diffraction signals\cite{noguchi00,beale10} could be accounted for by the distortions of the crystal structure local to the vanadium sites, \textit{i.e}. ATS scattering, without the need to invoke any $3d$ orbital order\cite{takahashi02,beale10}.

Figures \ref{resfig}a and \ref{resfig}b show the RXD spectra of the (011) and (010) Bragg forbidden reflections, respectively. Any differences observed between the RXD spectra measured at different temperatures reflect changes in the electronic environment of the vanadium ions. At both reflections we observed very little change between the 105 and 165~K data, either side of \tso. This agrees well with the results of the x-ray powder diffraction refinements, from which we surmise that the magnetic exchange interactions strongly couple to the lattice, causing a locking of the crystal structure, giving little change either side of \tso\ despite further cooling. Furthermore, there is a significant change in the RXD lineshape between 165 and 295~K, as one might expect upon a change of crystal symmetry at \too.

\section{\label{conc}Conclusions}

Through x-ray powder diffraction we have refined the lattice parameters of \smvo\ in the temperature range $12 < T < 300$ K. In doing so we have shown that the sample exists in a single phase below \tso. This contradicts the findings of Sage \etal\cite{sage06} We observed that the monoclinic distortion occurs and increases dramatically on cooling below \too, as expected. The monoclinic angle, $\alpha$, then locks into a given value at \tso, likely due to the magnetostriction occurring at \tso, which appears to stabilize the structural distortions within the crystal. Full crystal structure refinements again synchrotron x-ray powder diffraction data were performed, showing a substantial Jahn-Teller distortion of the vanadium - oxygen octahedra, compatible with G-type orbital order below \too\ and \tso. This strong structural distortion makes the existence of orbital dimerization, proposed to exist in other \rvo\ crystals, somewhat unlikely. Finally, we have probed the crystal field of the vanadium ions by employing resonant x-ray diffraction. Qualitatively, we showed little change in the crystal field upon cooling through \tso, and a more significant change upon the structural transition at \too. The structural behaviour of \smvo\ described herein is consistent with the physical trends observed in the \rvo\ series in general.

\begin{acknowledgments}
RDJ thanks M. Hayward and F. D. Romero for performing the thermogravimetric analysis. The authors would like to thank STFC and EPSRC for funding. We acknowledge the work performed on the EPSRC-funded XMaS beam line at the ESRF, directed by  M.J. Cooper and C. Lucas. We are grateful to the beam line team of S.D. Brown, P. Normile, D.F. Paul and P. Thompson for their invaluable assistance, and to S. Beaufoy and J. Kervin for additional support.
\end{acknowledgments}

\bibliography{smvo3bib.bib}

\end{document}